\documentclass[aip, pop, amsmath,amssymb,
 preprint,
]{revtex4-1}

\usepackage{graphicx}
\usepackage{dcolumn}
\usepackage{bm}

\usepackage[utf8]{inputenc}
\usepackage{etoolbox}

\usepackage{color}

\newcommand{\erf}{ \operatorname{erf} }
\newcommand{\wpl}{ \omega_\mathrm{p} }
\newcommand{\wplinv}{ \omega_\mathrm{p}^{-1} }
\newcommand{\lambdam}{ \lambda_{\alpha,\min} }
\newcommand{\rL}{ r_\mathrm{L0} }
\newcommand{\tha}{ \theta_\alpha }
\newcommand{\magza}{ \varkappa_{\alpha} }
\newcommand{\me}{{m_\mathrm{e}}}
\newcommand{\mi}{{m_\mathrm{i}}}

\makeatletter
\def\@email#1#2{%
 \endgroup
 \patchcmd{\titleblock@produce}
  {\frontmatter@RRAPformat}
  {\frontmatter@RRAPformat{\produce@RRAP{*#1\href{mailto:#2}{#2}}}\frontmatter@RRAPformat}
  {}{}
}%
\makeatother

\begin{document}

\title{Turbulent multicomponent magnetopause: Analytical description and kinetic simulation of complex distributed current sheets}

\author{A. A. Nechaev}
    \email{a.nechaev@ipfran.ru}
\author{M. A. Garasev}
\author{Vl. V. Kocharovsky}
\affiliation{A.V.~Gaponov-Grekhov Institute of Applied Physics, Russian Academy of Sciences, Nizhny Novgorod, 603950, Russia}

\date{\today}

\begin{abstract}
We carry out particle-in-cell simulations of complex current sheets of the family of analytically found Vlasov--Maxwell equilibria that model a collisionless magnetopause and allow for arbitrary energy distributions and countercurrents of various particle component. We find that, depending on the parameters, i)~a weak small-scale bending-type instability is always present at the sharp low-density plasma boundary of such a magnetopause, and ii)~the development of a small-scale Weibel-type instability inside a magnetopause is either excluded or possible. Both cases are studied and compared for two variants of the particle energy distribution~--- Maxwellian and Kappa. The long-term stability of the whole current sheet is observed either without or with saturated Weibel-type turbulence, which can be generated in a low-magnetized region between neighboring countercurrents. We point out the applicability of the developed model of distributed current sheets and the results of analysis of their small-scale instability for the description of physical phenomena in the magnetopauses of planets and late-type stars.
\end{abstract}

\maketitle

\section{Introduction. Models of~a~magnetopause with an~arbitrary energy distribution of~particles}
\label{sec:intro}

The article is devoted to modeling the inhomogeneous structure and kinetic processes in certain parts of magnetopauses where the flows of weakly collisional plasma as a whole are insignificant, but there are spatially separated flows of individual populations of charged particles, and, consequently, electric currents and the magnetic fields created by them. 
We shall restrict ourselves to the simplest plane-layered configurations, which are one-dimensionally inhomogeneous, localized on a finite section of the Cartesian $x$-axis and characterized by current densities $j_y$ flowing along the $y$-axis. These currents maintain a gradient in the only nonzero component of the magnetic field, $B_z$, directed along the $z$-axis and having given values, $B_z(x \to -\infty) > B_z(x \to +\infty)$, in the half-spaces to the left and right of the magnetopause (along the $x$-axis). Such smooth transitions between the half-spaces of quasihomogeneous magnetoactive plasma with different values of number density and magnetic field serve as approximations of certain parts of real magnetopauses, say, in the vicinity of the bow shock of planets and stars~\cite{Louarn2004, Dunlop2008, Norgren2018, Haaland2019, Burlaga2019}. Similar configurations can arise, for example, during the collision of strongly and weakly magnetized clouds of inhomogeneous stellar wind, or when the forming stellar wind flows around the upper part of a coronal loop of late-type stars. They can also be observed in laboratory experiments on plasma expansion into an external magnetic field~\cite{Shaikhislamov2009, Shaikhislamov2014}.

A necessary condition for the global equilibrium of a planar magnetopause is the constancy of the sum of the magnetic field pressure and the plasma pressure along the $x$-axis. However, in the presence of inhomogeneous currents and the anisotropy of particle velocity distributions, the fulfillment of this condition still leaves open the possibility of the development of various instabilities, primarily kinetic-type and especially small-scale ones. Analysis of the nature of turbulence arising in magnetopauses in collisionless plasma, where the mean free path of particles is much greater than the thickness of the transition layer, goes beyond the magnetohydrodynamic (MHD) approximation and requires the use of the Vlasov--Maxwell kinetic equations. 
This issue has been little studied to date, although it is important for describing real magnetopauses, where turbulence is omnipresent~\cite{Gurnett1979, Andre2001, Vaivads2004, Graham2019, Verscharen2019}.

Despite the simplicity of one-dimensional geometry, the structure of a planar magnetopause, even in the absence of turbulence, can be nontrivial, especially when taking into account the non-Maxwellian type of the actual particle energy distribution and the presence of spatially distributed currents, including countercurrents, of various particle components (fractions). In our papers a wide class of analytical stationary models of such complex multicomponent distributed magnetopauses was proposed~\cite{Nechaev23_JETPLen, Kocharovsky2021, Kocharovsky2019, Kocharovsky2016_UFN}. 
In this article, we present the first results of analytical estimates and a numerical study of the stability of these kinetic models, including the ones with a total thickness of the current sheet significantly exceeding the gyroradii of current-carrying particles. The results obtained show that kinetic instabilities occur in a wide range of parameters, but the developing small-scale turbulence quickly saturates at a fairly low level. Moreover, after its saturation, the global structure of the magnetopause does not qualitatively change during a long subsequent simulation, i.e., it is quasi-stationary and does not differ considerably from the initial current sheet found analytically without turbulence.

The global large-scale stability of the complex distributed current sheets considered here is no surprise due to the absence of a magnetic field's null point inside them and the presence of an external magnetic field outside. 
In this respect, they qualitatively differ from the Harris-type sheets widely applied to model current sheets in the stellar wind and planetary magnetotails~\cite{Harris1962, Allanson2017, Neukirch2020, Zelenyi2011, Tsareva2023, Lotekar2022, Vasko2022, Vasko2024}. These models include a region where the magnetic field changes direction near a null point and, as such, are prone to various large-scale MHD and small-scale kinetic instabilities (including the Weibel one) associated with the reconnection of magnetic field lines (see, e.g., Refs.~\onlinecite{Sitnov2002, Ricci2004, Karimabadi2004, Sitnov2017, Akcay2016, Jain2017, Sironi2014}). 
However, to the best of our knowledge, previously conducted particle-in-cell (PIC) simulations of such sheets indicate that the development of various instabilities, even if saturated, results in strong turbulence and cardinal change of their original global structure.

The magnetopause models proposed and studied here include asymmetric multipeak current sheets with arbitrary energy distributions of particles. They can be useful for comparative study of the complex current sheets in the outer magnetospheres of planets or exoplanets and late-type stars, in particular, for interpreting observations of the Earth's magnetopause (see, e.g., Refs.~\onlinecite{Burch2016, Norgren2018, Louarn2004, Dunlop2008, Haaland2019, Shuster2021}). These models can shed light on the connection between the spatial profiles of quasi-static currents, the anisotropic velocity distributions of particles, and the features of small-scale turbulence in the magnetopause region. It is also worth analyzing how the development and saturation of turbulence at the boundary and inside the complex current sheet depend on its structure and parameters. The problem is especially challenging if a magnetopause has an extended region with a relatively weak magnetic field and sufficiently strong anisotropy of particle velocity distributions.

In this work, we attempt to answer these questions on the basis of specific, yet quite representative, examples of two-component (Sect.~\ref{sec:2cMaxw}) and four-component (Sects.~\ref{sec:simul}~B, C) current sheets, found analytically for an arbitrary energy distribution of particles~\cite{Nechaev23_JETPLen}. For definiteness, we limit ourselves to two such distributions, Maxwellian and Kappa (see Sect.~\ref{sec:theory:models}), and focus on the Weibel-type kinetic instability using the analytical self-consistent stationary solutions to the Vlasov--Maxwell equations as initial conditions in numerical PIC-simulations with the code EPOCH~\cite{Arber2015}. Previously, a consistent analysis of both the stability of the complex current sheets mentioned above and the development of the Weibel turbulence in the magnetopause was absent.

The main purpose of this work is to show the typical structure and parameters of complex current sheets in which the Weibel-type instability is possible despite (a)~the presence of a self-consistent magnetic field, (b)~not very high anisotropy of the particle velocity distribution, and (c)~a limited region of sufficiently dense nonequilibrium plasma in the magnetopause. We present some illustrative examples of numerical simulations of the transient evolution of the proposed current sheets both in the absence and in the presence of the Weibel-type instability. We observe this instability only in the specific cases (not all) of the four-component magnetopause with countercurrents and a wide region of a weak self-consistent magnetic field inside. In all such cases, the small-scale instability is saturated at a fairly low level and does not considerably change the overall structure of the initial current sheet. 

As for the instability at the sharp boundary of a current sheet with strong gradients of the magnetic field and plasma density, which resembles the bending instability discussed in literature~\cite{Bertin1985, Bulanov1996, Califano1999}, it is typical for our magnetopause models. This instability occurs for all examples considered in the paper, saturates at a quite low level and results in a weak turbulence involving the plasma components with different energies, in particular, the low-energy background particles responsible for the local neutrality of the magnetopause (see Sect.~\ref{sec:theory:models}). A detailed analysis of this bending-type instability and the corresponding persistent fluctuations of the magnetopause boundary is beyond the scope of this work.

\section{Analytical description of current sheets and stability estimates}
\label{sec:theory}

\subsection{Current sheet models}
\label{sec:theory:models}

Let us first recall some basic equations used in the suggested models of boundary current sheets following our work~\cite{Nechaev23_JETPLen}.

The distribution function for each plasma component is chosen as an isotropic function of the absolute value of the momentum $p = |\vec{p}|$, i.e., the particle energy, multiplied by the Heaviside step function of one of the projections of the generalized momentum:
\begin{eqnarray}
    f_\mathrm{e} (x, p, p_y) & = & N_{\mathrm{e}1} F_{\mathrm{e}1}(p) \, H\! \left( -p_y + \frac{e}{c} [ A_y(x) - A_{\mathrm{e}1} ] \right)
    + N_{\mathrm{e}2} F_{\mathrm{e}2}(p) \, H\! \left( p_y - \frac{e}{c} [ A_y(x) - A_{\mathrm{e}2} ] \right)
        \nonumber \\
    && + n_{\mathrm{e}0}(x) F_{\mathrm{e}0}(p) ,
\label{eqDistribFunce4comp} \\
    f_\mathrm{i} (x, p, p_y) & = & N_{\mathrm{i}1} F_{\mathrm{i}1}(p) \, H\! \left( p_y + \frac{e}{c} [ A_y(x) - A_{\mathrm{i}1} ] \right)
    + N_{\mathrm{i}2} F_{\mathrm{i}2}(p) \, H\! \left( -p_y - \frac{e}{c} [ A_y(x) - A_{\mathrm{i}2} ] \right)
    \nonumber \\
    && + n_{\mathrm{i}0}(x) F_{\mathrm{i}0}(p) .
\label{eqDistribFunci4comp}
\end{eqnarray}
Here the plasma components are denoted by the indices $\alpha = \mathrm{e},\mathrm{i}$, corresponding to electrons or ions respectively, and $s = 1,2$, which serves to distinguish countercurrents of particles: $s = 1$ determines the current directed along the Cartesian coordinate axis $y$, $s = 2$ is against it.
For each component, the constant $N_{\alpha s}$ and the isotropic factor $F_{\alpha s}(p)$ characterize the number density and energy distribution of particles on one side of the sheet, in the half-space $(-1)^{s+1 } x \to +\infty$, where this component is isotropic.
$H (\xi)$ is the Heaviside function such that $H (\xi > 0) = 1$ and $H (\xi <= 0) = 0$.
The vector potential $\vec{A}$ defining the magnetic field $B_z (x)$ of the current sheet has a single nontrivial component $A_y$, so that $B_z = \mathrm{d} A_y / \mathrm{d} x$.
Constant shifts $A_{\alpha s} = \mathrm{const}$ can be chosen different for different plasma components.
For simplicity, ions are assumed to be single-charged, $e$ is the elementary charge, $c$ is the speed of light in vacuum.
Finally, the velocity isotropic terms with indices~$0$ in each distribution function (\ref{eqDistribFunce4comp}), (\ref{eqDistribFunci4comp})  do not contribute to the current and are introduced in order to guarantee the electrical neutrality of the current sheets. In the simulations (Sect.~\ref{sec:simul}) we assume they are Maxwellian.

An analysis of the general properties of the stationary current sheets specified by the distribution functions (\ref{eqDistribFunce4comp}) and (\ref{eqDistribFunci4comp}) with arbitrary energy profile $F_{\alpha s}(p)$ was carried out in~\cite{Nechaev23_JETPLen}.
In the numerical simulations presented in Section~\ref{sec:simul} we are using only Maxwellian and Kappa distributions, so let us provide expressions for these two cases.

For a current sheet with a Maxwellian distribution of the form
\begin{equation}
    F_{\alpha s} (p) = \left( 2 \pi m_\alpha T_{\alpha s} \right)^{-3/2} \exp \left( - \frac{p^2}{p_{\alpha s}^2} \right)
\label{eqDistribFuncMaxw}
\end{equation}
it is easy to obtain the following expressions for the profiles of the current density $j_y(x)$ and the number density $n_{\alpha s}(x)$ of particles by simply integrating the functions (\ref{eqDistribFunce4comp}) and (\ref{eqDistribFunci4comp}) over the momenta:
\begin{equation}
    j_y(A_y) = \sum\limits_{\alpha, s} j_{\alpha s} = \sum\limits_{\alpha, s} \frac{(-1)^{s+1}}{2 \pi^{1/2}} e N_{\alpha s} \frac{p_{\alpha s}}{m_\alpha} \exp\left( - a_{\alpha s}^2 \right) ,
\label{eqjMaxw}
\end{equation}
\begin{equation}
    n_{\alpha}(A_y) = \sum\limits_s n_{\alpha s} = \sum\limits_s \frac{1}{2} N_{\alpha s} \left[ 1 + (-1)^{s+1} \erf \left( a_{\alpha s} \right) \right] ,
\label{eqconcMaxw}
\end{equation}
where we define the dimensionless vector potential $a_{\alpha s} (x) = [A_y(x) - A_{\alpha s}] \, e / (p_{\alpha s} c)$, the thermal particle momentum $p_{\alpha s} = \left( 2 m_\alpha T_{\alpha s} \right)^{1/2}$, and the effective temperature $T_{\alpha s} = \langle p^2 \rangle_{\alpha s} / (3 m_{\alpha})$ of the $\alpha s$ plasma component in the region of its isotropy, at $(-1)^{s+1} A_y \to +\infty$. We also denote $\langle\dots\rangle_{\alpha s} = \int\! (\dots) f_{\alpha s} \, d^3\vec{p} \,/ n_{\alpha s}$ and $\erf (\xi) = 2 \pi^{-1 /2} \int_0^\xi \exp \left( -t^2 \right) d t$.
As can be shown~\cite{Nechaev23_JETPLen} for arbitrary energy distribution $F_{\alpha s} (p)$, each component's current density, $j_{\alpha s}$, has the same sign for all values of $A_y$, achieves a single extremum at $a_{\alpha s} = 0$ and tends to $0$ as $A_y \to \pm\infty$; number densities $n_{\alpha s}(a_{\alpha s})$ are monotonic functions of their arguments.

The magnetic field of the sheet can be found from the pressure balance condition:
\begin{equation}
    B_z (A_y) = (8 \pi)^{1/2} \left[ P_0 - P_{xx}(A_y) \right]^{1/2} ,
\label{eqBgeneral}
\end{equation}
\begin{equation}
    P_{xx}(A_y) = \sum\limits_{\alpha, s} P_{\alpha s}= \sum\limits_{\alpha, s} (-1)^{s+1} \frac{1}{2} N_{\alpha s} T_{\alpha s}
    \left( \erf\, a_{\alpha s}
    - \erf \left.a_{\alpha s}\right|_{x = 0} \right) ,
\label{eqPxxMaxw}
\end{equation}
where $P_{xx}$ is the $xx$-component of the plasma pressure tensor formed by the sum of each particle component, $P_{\alpha s} = m_\alpha^{-1} n_{\alpha s} \langle p_x^2 \rangle_{\alpha s} + \textrm{const}$, with negative constants ensuring $P_{xx}(0) = 0$.
In all presented current sheet models we choose the parameter $P_0$ such that the expression in square brackets is positive. Then the magnetic field component $B_z$ does not change sign throughout the current sheet, varying from one constant on the left side of the sheet to another constant on the right side. For convenience, the direction of the $z$ axis is chosen so that this sign is positive.

Finally, the transition from the vector potential to the coordinate is defined by the expression
\begin{equation}
    x(A_y) = \int\limits_0^{A_y} \! \frac{d A'}{B_z(A')} \; ,
\label{eqxAy}
\end{equation}
where the limits of integration are chosen so that at the point $x = 0$, which we call the center of the sheet, the vector potential $A_y = 0$. Since the magnetic field component $B_z$ is positive everywhere, the function $x(A_y)$ is monotonic and invertible, so that the destination of the limits $x \to \pm \infty$ requires unlimited values $A_y \to \pm \infty$, respectively.

Expressions~(\ref{eqjMaxw})--(\ref{eqxAy}) define a model of an equilibrium current sheet in collisionless plasma representing a distributed magnetopause, i.e., the transition between two plasma half-spaces with different number densities and temperatures of particles, and different magnitudes of the transverse magnetic field.

Similarly, for a current sheet with a Kappa energy distribution
\begin{equation}
    F_{\alpha s} (p) = \frac{ M_\kappa}{\pi^{3/2} p_{\alpha s}^3}
    \left( 1 + \frac{p^2}{\left( \kappa-3/2 \right) p_{\alpha s}^2} \right)^{-\kappa-1} ,
\label{eqDistribFuncKappa}
\end{equation}
where $\kappa > 3/2$, $p_{\alpha s} = \left( 2 m_\alpha T_{\alpha s} \right)^{1/2}$, and the normalization factor is
\[
    M_\kappa = \left( \kappa-\frac{3}{2} \right)^{-3/2} \frac{\Gamma(\kappa+1)}{\Gamma(\kappa-1/2)} ,
\]
it is easy to obtain expressions~\cite{Kocharovsky2021} for the current and number densities:
\begin{eqnarray}
    j_{\alpha s}(A_y) & = & \frac{(-1)^{s+1} }{2 \pi^{1/2} } e N_{\alpha s}
    \frac{p_{\alpha s}}{m_\alpha}  M_\kappa
    \frac{ \left( \kappa-3/2 \right)^2 }{\kappa (\kappa -1)}
    \left( 1 + \frac{a_{\alpha s}^2}{ \kappa-3/2 } \right)^{-\kappa + 1} ,
\label{eqjKappa}
\end{eqnarray}
\begin{equation}
    n_{\alpha s}(A_y) = \frac{N_{\alpha s}}{2} + \frac{(-1)^{s+1}}{\pi^{1/2}} N_{\alpha s}
    M_\kappa
    \frac{\kappa-3/2}{\kappa}
    \, a_{\alpha s} \, G_\kappa \! \left( -\frac{a_{\alpha s}^2}{\kappa - 3/2} \right) ,
\label{eqconcKappa}
\end{equation}
where we use the Gaussian hypergeometric function $G_\kappa (\xi) = {}_2 F_1 \! \left( \frac{1}{2}, \kappa; \frac{3}{2}; \xi \right)$. The magnetic field can be found from Eq.~(\ref{eqBgeneral}) with substitution
\begin{equation}
    P_{xx} (A_y) 
    = \sum\limits_{\alpha, s} \frac{(-1)^{s+1} }{\pi^{1/2}} N_{\alpha s} T_{\alpha s}
    M_\kappa
    \frac{ \left( \kappa-3/2 \right)^2 }{\kappa (\kappa -1)}
    \, a_{\alpha s} \, G_{\kappa-1} \! \left( -\frac{a_{\alpha s}^2}{\kappa - 3/2} \right) 
    +\mathrm{const},
\label{eqPxxKappa}
\end{equation}
where the constant is chosen so that $P_{xx}(A_y = 0) = 0$. 
For $\kappa \to \infty$, Eqs.~(\ref{eqjKappa})--(\ref{eqPxxKappa}) turn into Eqs.~(\ref{eqjMaxw}),~(\ref{eqconcMaxw}) and (\ref{eqPxxMaxw}), respectively.
For the half-integer $\kappa = K/2$, $K \in \mathbb{N}$, function $G_\kappa$ can be easily recast via elementary functions~\cite{Prudnikov1990}. In Section~\ref{sec:4cKappa} we will bear in mind typical power laws of the energetic-particle distribution in the Earth's magnetopause and set the parameter $\kappa = 3$, for which
\[
    G_3 \left( -\xi^2 \right) = \frac{3 \xi^2 + 5}{8 \left(1 + \xi^2\right)^2} + \frac{3 \arctan \xi}{8 \xi} ,\quad 
    G_2 \left( -\xi^2 \right) = \frac{1}{2 \left(1 + \xi^2\right)} + \frac{\arctan \xi}{2 \xi} .
\]

The parameter $\tau_\alpha = 1 - \left( \langle p_y^2 \rangle - \langle p_y \rangle^2 \right) / \langle p_x^2 \rangle$, characterizing the anisotropy of particles of the sort $\alpha$, can be found for an arbitrary energy distribution, including the Maxwellian one~\cite{Kocharovsky2019}:
\begin{equation}
    \tau_{\alpha} (A_y) =
    \frac{ \sum\limits_{s} p_{\alpha s} a_{\alpha s} j_{\alpha s} 
    + e^{-1} m_\alpha n_\alpha^{-1} j_{\alpha}^2
    }{
    e m_\alpha^{-1} \sum\limits_{s} n_{\alpha s} \langle p_x^2 \rangle_{\alpha s}
    } \, ,
\label{eqAnisGen}
\end{equation}
where $j_\alpha = \sum_{s} j_{\alpha s}$.
For each individual plasma component ($\alpha s$), at $(-1)^{s+1} a_{\alpha s} \to +\infty$ its distribution function is isotropic, $\tau \to 0$, while at $( -1)^{s+1}a_{\alpha s} \to -\infty$, where the particle number density vanishes, $n_{\alpha s} \to 0$, its anisotropy parameter reaches maximum, $\tau \to 1$, since there exist only particles with large $y$ projections of momentum.

The profiles of the magnetic field, current density, and number density for the Maxwellian and different Kappa energy distributions are qualitatively similar, but can have significantly different localization scales~\cite{Kocharovsky2021} for small $\kappa$ values close to the $3/2$ limit. In all cases, we assume that the whole magnetopause current sheet has a finite size along $x$-axis with a maximum fixed value of magnetic field, $B_z (x \to -\infty)$, at the left side and a smaller fixed value of magnetic field, $B_z (x \to +\infty)$, at the right side. In such a magnetopause the magnetic-field profile can be nonmonotonic due to presence of the separate currents of different particle components.

\subsection{Analysis of stability}
\label{sec:theory:stability}

Let us now proceed to the analysis of the properties of the electron Weibel instability (the growth rate of the ion instability is very small due to large ion masses) on the basis of the well-known linear theory for the bi-Maxwellian particle distribution, which we use for qualitative order-of-magnitude estimates for both the case of truncated Maxwellian distributions~(\ref{eqDistribFunce4comp})--(\ref{eqDistribFunci4comp}) with substitution~(\ref{eqDistribFuncMaxw}), and for the case of Kappa distributions~(\ref{eqDistribFuncKappa}). (Generally, the properties of instability can significantly depend on the type of the distribution function; see, e.g., Refs.~\onlinecite{Kocharovsky2016_UFN, Silva2021, Kuznetsov2022}.)

Consider a simple situation where a single anisotropic electron component, $s = 1$, is present. As already mentioned, the quantity $\tau_\mathrm{e}$ given by~(\ref{eqAnisGen}) is positive everywhere, and the distribution of electrons in velocity space has the form of a ''disk'', since 
\begin{equation}
    \tilde{T}_x = \tilde{T}_z \geq \tilde{T}_y, \quad \vec{B}(x) \parallel Oz ,
\label{eq:geometry}
\end{equation}
where $\tilde{T}_j = ( \langle p_j^2 \rangle - \langle p_j \rangle^2 ) / m_\mathrm{e}$ is the effective temperature along a certain axis ($j = x,y,z$) and the index $\alpha = \mathrm{e}$ is omitted for brevity.

As was said, for rough estimates we assume the electron distribution to be bi-Maxwellian and use the detailed linear theory~\cite{Weibel1959, Davidson1989, Vagin2014, Borodachev2017}, ignoring the presence of the magnetic field of the current sheet.
It is well-known~\cite{Vagin2014} that for a bi-Maxwellian distribution two aperiodic modes are unstable, and perturbations with wave vectors oriented in the direction of the smallest dispersion of particle velocities have the highest growth rates.
From those, only perturbations with a wavelength less than some definite value $\lambdam$ can grow. In the approximation of a small anisotropy parameter, the perturbations on the wavelength of $\lambda_\mathrm{e} = \sqrt{3} \lambdam$ have the highest growth rate of $\Gamma$, where
\begin{equation}
    \lambdam \approx \frac{2 \pi c}{\wpl}
        \tha^{-1/2},
\label{eqLambdaMin}
\end{equation}
\begin{equation}
    \Gamma \approx \wpl
        \left( \frac{n_\mathrm{e} \tilde{T}_x}{N_\mathrm{e} \me c^2} \right)^{1/2}
        \frac{1}{\pi^{1/2}} \left( \frac{2 \tau_\mathrm{e}}{3} \right)^{3/2} ,
\label{eqGammaMax}
\end{equation}
$\omega_\mathrm{p} = (4 \pi e^2 N_\mathrm{e} / \me)^{1/2}$ is the plasma frequency, and $\tha = (n_\mathrm{e} / N_\mathrm{e}) \, \tau_\mathrm{e} (1-\tau_\mathrm{e})^{-1}$ is the effective anisotropy that determines the instability properties.

In the simplest case of a single plasma component, as seen from Eqs.~(\ref{eqconcMaxw}) and (\ref{eqAnisGen}), the effective anisotropy $\tha$ is positive only in the vicinity of the point $a_\mathrm{e} = 0$ and almost vanishes already at $|a_\mathrm{e}| > 2$, being described as $\propto |a_\mathrm{e}| \exp (-a_\mathrm{e}^2)$ in the limits $a_\mathrm{e} \to \pm \infty$.
Its maximum is at $a_\mathrm{e} \approx -0.1$, so we can estimate: $\max \tha \approx \tha (0) = (\pi - 2)^{-1}$.
(Since $\tha < 1$, the approximation of plasma dispersion function used in the Eqs.~(\ref{eqLambdaMin}),~(\ref{eqGammaMax}) is adequate.)

Let us now compare the electron gyroradius $r_{\mathrm{L}\alpha} = (2 \me \tilde{T}_x)^{1/2} c (e B_z)^{-1}$, where $\tilde{T}_{x,z} (x) \equiv T_\mathrm{e}$ for the Maxwellian energy distribution~(\ref{eqDistribFuncMaxw}), with the half of the minimal wavelength of unstable pertubations, $\lambdam / 2$.
Using Eq.~(\ref{eqLambdaMin}), we obtain the following demagnetization parameter
\begin{eqnarray}
    \magza \equiv \left( \frac{2 \, r_{\mathrm{L}\alpha}}{\lambdam} \right)^2 =
        \frac{\tha}{\pi^2} \, \frac{8 \pi N_\mathrm{e} T_\mathrm{e}} {B_z^2} .
\label{eqCriterium}
\end{eqnarray}
As follows from the well-known saturation criterion for the Weibel instability (see, e.g., Refs.~\onlinecite{Bret2013, Kocharovsky2016_UFN}), for the geometry under consideration, the inequality $\magza < 1$ excludes the instability of pertubations with wave vectors directed along the $y$-axis. At $a_{\mathrm{e}} = 0$, where the effective anisotropy $\tha$ is nearly maximum, we have $B_z^2(0) = 8 \pi P_0$ and the parameter $\magza (0) \leq 2 \pi^{-2} \tha(0) < 0.2$, which follows from the condition $P_0 \geq \max P_{xx}(A_y) = N_\mathrm{e} T_\mathrm{e} / 2$ assumed in Eq.~(\ref{eqBgeneral}) for all constructed models.
Note that even in the limiting case, $2 P_0 = N_\mathrm{e} T_\mathrm{e}$, we have $\magza(|a_{\mathrm{e}}| \leq 2) \lesssim 0.9$, although outside of this region, for $a_{\mathrm{e}} > 2$, the magnetic field is vanishing, $B_z \to 0$, and formally $\magza \to \infty$. However, even the maximum growth rate, given by Eq.~(\ref{eqGammaMax}), rapidly goes to zero here, $\Gamma \to 0$ (and $\lambda_\mathrm{e} \to \infty$), and, hence, there is no instability.
Thus, in a simplest model, particles are always magnetized in the sheet's field regardless of the $N_{\mathrm{e}}$ and $T_{\mathrm{e}}$ values, so the Weibel instability is inhibited even for the minimal wavelengths. 

In case there are multiple plasma components, but all of them form currents in the same direction (e.g., for electrons the distribution $f_\mathrm{e} = \sum_{s} N_{\mathrm{e} s} F_{\mathrm{e} s}(p) \, H ( -p_y + \tilde{A}_{\mathrm{e} s})$ means that all $j_{\mathrm{e} s} > 0$), the function $P_{xx} (A_y)$ is again monotonic, so we can presume that $\varkappa_\alpha \lesssim 2 \pi^{-2} \theta_\alpha$ in the finite region of instability.
If all the shifts $A_{\alpha s}$ are zero, $A_{\alpha s} = 0$, or equal to each other, we have the inequality~\cite{Kocharovsky2019} $\tau_\alpha(0) \leq 2 / \pi$ leading to $\theta_\alpha(0) \leq (\pi - 2)^{-1}$. However, it is challenging to find how the value $\max \theta_\alpha$ is close to $\theta_\alpha(0)$ and whether the inequality $\max \varkappa_\mathrm{e} < 1$ is fulfilled for all sets of plasma parameters, so that the inhibition of Weibel-type instability is questionable.

When countercurrents are present, the condition $2 P_0 \geq \max \sum_{s} (-1)^{s+1} N_{\alpha s} T_{\alpha s} \, \erf (a_{\alpha s})$ does not infer any restrictions on the demagnetization parameter $\varkappa_\mathrm{e}$. This fact is used in Section~\ref{sec:simul} to construct a Weibel-unstable current sheets with small-scale turbulence inside.
Moreover, a certain choice of plasma parameters can provide the condition $\tau_\alpha < 0$ in some range of the vector-potential component $A_y$. Nevertheless, by increasing the value $P_0$ one can ensure the inequality $\varkappa_\mathrm{e} < 1$ and hence the absence of the Weibel-type instability inside a current sheet of the presented class.

Inhibition of the Weibel instability by a sufficiently strong magnetic field parallel to the axis of the highest temperature, $\vec{B} \parallel Oz$, is also expected from linear theory. Indeed, as follows from the dispersion equation for transverse modes in a homogeneous bi-Maxwellian plasma with temperatures $\tilde{T}_z > \tilde{T}_x = \tilde{T}_y$ in the presence of an external homogeneous magnetic field directed along z-axis, the inequality $\magza < 1$ approximately coincides with the condition of collapse of the interval of unstable wavenumbers~\cite{Emelyanov2024}. 
Note, however, that such a magnetized anisotropic plasma can still fuel the oblique firehose instability (OFI), which corresponds to the extraordinary-wave branch of the dispersion relation.
It is well-studied in the literature (see.,~e.g., Refs.~\onlinecite{Cozzani2023, Lopez2019, Camporeale2008, Gary2003}) for bi-Maxwellian and bi-Kappa particle distributions with effective temperatures $\tilde{T}_z > \tilde{T}_x = \tilde{T}_y$, mainly for the parameters of the near-Earth solar wind plasma. 
This instability is also aperiodic (nonpropagating) and has a positive growth rate for the wave vectors lying in the cone of intermediate angles to the magnetic field around some optimal angle (approximately 70{\textdegree} for typical solar wind parameters~\cite{Cozzani2023}). For the orthogonal wave vectors, the growth rate is zero in the case of a bi-Maxwellian energy distribution and is much smaller than maximum in the case of a bi-Kappa distribution~\cite{Lopez2019}. 

In the current sheets considered here the momentum distribution is not axially symmetric with respect to the magnetic field~(\ref{eq:geometry}). The analytical dispersion relation in this case is missing, so let us again use for estimates the linear theory for a bi-Maxwellian plasma~\cite{Gary2003}. 
In the simplest one-component models, the parallel plasma beta is small near the center of the sheet, $\beta_\parallel = 8 \pi n_\mathrm{e} \tilde{T}_z / B_z^2 \leq 1$, hence OFI is precluded.  
In the sheets with countercurrents, we can simultaneously achieve both inequalities, $\magza < 1$ and $\beta_\parallel = \magza (\pi^2 / \tha) (n_\mathrm{e} / N_\mathrm{e}) > 1$, by adjusting the plasma parameters, so that OFI can be observed alone, while the Weibel instability is still inhibited. (Note that OFI growth rate is smaller than the electron gyrofrequency and thus smaller than the Weibel growth rate $\Gamma$ given by Eq.~(\ref{eqGammaMax}).)
To confirm these conclusions one should use full 3D simulations, since the wave vectors in the plane $yz$ cannot be neglected. At present this is beyond the reach of our computational resources.

\section{Numerical modeling}
\label{sec:simul}

To verify the analytical estimates, we performed numerical kinetic PIC-simulations using the code EPOCH~\cite{Arber2015}.
Calculations were performed for two models of current sheets to demonstrate both the stability of simple models without countercurrents and the possibility of constructing a sheet that is unstable with respect to the Weibel perturbations.
In all simulations, a model ion mass of $\mi = 18\, \me$ was used, which reduced the total width of the sheet, proportional to ion gyroradius, and hence the required size of the simulation domain. In addition, light ions respond more quickly to the force exerted by the electric field, which arises due to incomplete compensation of the charge in the central part of the sheet and disrupts its stationarity. In current sheets with ions of realistic mass, the influence of this factor will be weaker. For simplicity, we use two-dimensional simulations (2D3V), assuming homogeneity along the $z$-axis and taking into account the kinetics of all three components of the particle velocity $\vec{v}$. This approach is correct for the most unstable pertubations with the wave vectors $\vec{k}$ directed along y-axis.

\subsection{Two-component stable current sheet with Maxwellian energy distributions of particles}
\label{sec:2cMaxw}

\begin{figure}[b]
\centering
    \includegraphics[width=0.9\linewidth]{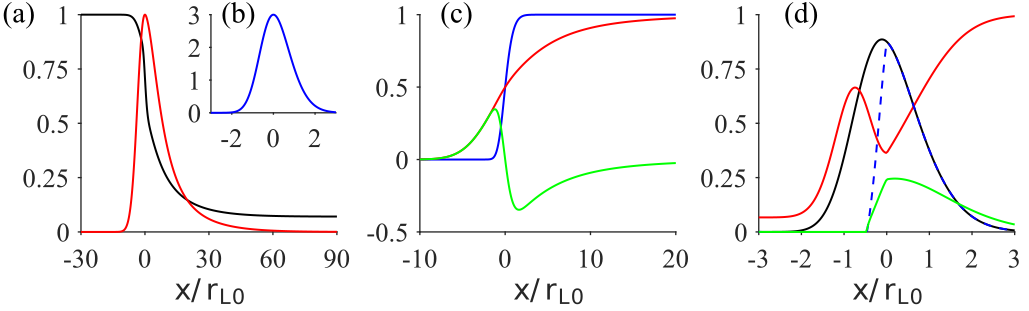}
	\caption{
        Structure of the two-component current sheet at the initial moment of time:
        (a)~black~--- initial profile of the magnetic field $B_{z0}$ in the simulation domain $-30 < x / \rL < 90$, normalized to its maximum $B_{z0} (-\infty)$
        $ = 2 \pi^{1/2} ( N_\mathrm{e1} T_\mathrm{e1} + N_\mathrm{i1} T_\mathrm{i1} )^{1/2} (\tilde{P}_0 + 1)^{1/2}$; 
        red~--- ion current density $j_\mathrm{i1}$ in projection on the axis~$y$, normalized to its maximum $j_\mathrm{i1}(0) = (2 \pi^{1/2})^{-1} e N_\mathrm{i1} p_\mathrm{i1} / \mi$;
        (b)~the same projection of electron current density $j_\mathrm{e1}$, normalized to $j_\mathrm{i1}(0)$;
        (c)~red~--- normalized number density profile of current-carrying ions $n_\mathrm{i1}/N_\mathrm{e1}$, blue~--- same for electrons $n_\mathrm{e1}/N_\mathrm{e1}$, green~--- their difference $(n_\mathrm{i1} - n_\mathrm{e1})/N_\mathrm{e1}$, proportional to their charge density;
        (d)~black~--- effective anisotropy of current-carrying electrons $\theta_\mathrm{e1}$, red~--- effective temperature $\tilde{T}_y$ of all electrons including background, blue~--- their joint effective anisotropy $\theta_\mathrm{e}$, green~--- their joint demagnetization parameter~$\varkappa_\mathrm{e}^{1/2}$.
	}
	\label{fig1}
\end{figure}

Let us consider a sheet formed by one electron and one ion components with Maxwellian energy distributions.
In the isotropic region, the temperatures of electrons and ions are $T_\mathrm{e1} = 150$~eV and $T_\mathrm{i1} = 300$~eV, respectively, the number densities are equal to $N_\mathrm{e1,i1} = 10^6$~cm$^{-3}$. Constant shifts are absent: $A_\mathrm{e1,i1} = 0$. 
We choose the magnetic field~(\ref{eqBgeneral}) to be very close to zero on the right-hand side of the sheet: $B_z (+\infty) = 2 \pi^{1/2} ( N_\mathrm{e1}T_\mathrm{e1} + N_\mathrm{i1}T_\mathrm{i1} )^{1/2} ( \tilde{P}_0 - 1 )^{1/2}$, where $\tilde{P}_0 = 1.01$. Here and below, in the multicomponent cases, we introduce the dimensionless parameter $\tilde{P}_0 = 2 P_0 / ( \sum_{\alpha, s} N_{\alpha s} T_{\alpha s} )$. 
Background electrons and ions, neutralizing the charge in the central part of the sheet, have Maxwellian isotropic velocity distributions with the temperature $T_\mathrm{e0,i0} = 10$~eV and number density profiles $n_\mathrm{e0} (x) = \max \{ 0, n_\mathrm{i1} - n_\mathrm{e1} \}$, $n_\mathrm{i0} (x) = \max \{ 0, n_\mathrm{e1} - n_\mathrm{i1}\}$.
Profiles of the sheet's magnetic field, current densities, and particle number densities at the initial moment of time are shown in Fig.~\ref{fig1}.

The gyroradius of the thermal electron in the center of the sheet is equal to $\rL \approx 430$~cm, and the width of the sheet does not exceed $70 \rL$. The effective anisotropy parameter $\tha$ (see Fig.~\ref{fig1}) has a maximum at $x \approx -0.1 \, \rL$ of the order of $\tha \approx 0.9$, which, according to~(\ref{eqLambdaMin}), in the absence of a magnetic field would correspond to the wavelength of the fastest growing Weibel perturbations $\sqrt{3} \lambdam \approx 12 \, c/\wpl$. (Background electrons with the chosen density profile $n_\mathrm{e0} (x)$ reduce the joint anisotropy of electrons on the left slope of the sheet, where they dominate in number, but do not significantly change the maximum of the anisotropy parameter $\tha$.)
Taking these estimates into account, the size of the two-dimensional simulation region is chosen to be equal to $L_x = 100 \, c/\wpl \approx 120 \, \rL$ and $L_y = 100 \, c/\wpl \approx 15 \, \lambdam$ along the axes~$x$ and~$y$, respectively. The boundary conditions at $x = -L_x / 4$ and $x = 3 L_x / 4$ are reflective for particles and absorbing for fields, the boundary conditions at $y = 0$ and $y = L_y$ are periodic.
The region was divided into $1250 \times 1250$ cells, the current-carrying plasma components comprise approximately $75 \times 10^6$ macroparticles, and background components comprise approximately $25 \times 10^6$ ones. 

According to analytical estimates in Section~\ref{sec:theory}, in a sheet without countercurrents, one should not expect the development of the Weibel instability.
Indeed, as an exact calculation shows, the maximum demagnetization parameter~(\ref{eqCriterium}) of all electrons in the sheet is $\varkappa_\mathrm{e} (x \approx 0.2\, \rL) \approx 0.06$.
If, for the same plasma parameters, the electrons were not magnetized by the current sheet's field, the growth time of the Weibel perturbations (their inverse maximum growth rate $\Gamma^{-1}$) would be of the order of $370\,\wplinv$.
(The value is obtained from the relation~(\ref{eqGammaMax}) substituting the effective electron temperature $\tilde{T}_y \approx 0.4 \, T_\mathrm{e1}$, see Fig.~\ref{fig1}.)

\begin{figure}[t]
\centering
    \includegraphics[width=0.8\linewidth]{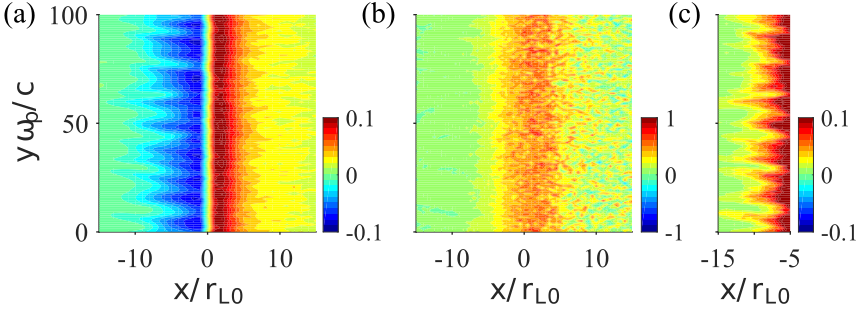}
	\caption{
        Two-component Maxwellian current sheet described by the following two-dimensional distributions at the moment $t = 4900 \, \wplinv$:
        (a)~ the magnetic field perturbation, $B_z(x,y) - B_{z0}(x,y)$, normalized to $\max B_{z0}$;
        (b)~$y$-component of the current density of all particles, $j_y(x,y)$, normalized to $j_\mathrm{e1}(t = 0, x = 0) = (2 \pi^{1/2})^{-1} e N_\mathrm{e1} p_\mathrm{e1} / \me$;
        (c)~number density of all electrons, $(n_\mathrm{e1} + n_\mathrm{e0}) / N_\mathrm{e1}$.
	}
	\label{fig2}
\end{figure}

The simulation has been carried out until $t = 17500 \, \wplinv$ and does not show the Weibel-type instability.
Nevertheless, already at times of the order of $800 \, \wplinv$, in the distribution of the magnetic field $B_z$ and the number density of all electrons $n_\mathrm{e1} + n_\mathrm{e0}$ a modulation along the axis~$y$ appears at the left side of the sheet, with its magnitude reaching a maximum at the moment $t \approx 4900 \, \wplinv$ shown in Fig.~\ref{fig2}. 
This instability is not of a Weibel type and develops mainly at the sharp magnetopause boundary, which is close to the positions of maximum gradients of the plasma density and magnetic field. It may be called a ''plasma--vacuum'' boundary as well, since in the region $x / \rL < -1$, where the magnetic field is high, the number density of particles is insignificant, so that the anisotropy of electrons vanishes and the Weibel instability should be inhibited. 
Indeed, here the current density component $j_z$ contains only numerical noise and the number density is modulated similarly to the main component $j_y$ that forms the sheet. In the central part and at the high-density right edge of the magnetopause (Fig.~\ref{fig2}(b)), the latter current density component is almost unperturbed.

Such an instability, sometimes called
the bending instability, is typical for the layers of an inhomogeneous magnetoactive plasma with strong density gradients, including the sharp free boundaries with vacuum, and a constant sum of the magnetic field pressure and the particle kinetic pressure (cf., e.g., Refs.~\cite{Bertin1985, Bulanov1996, Califano1999}). 
This bending-type instability is observed in all our kinetic simulations of the magnetopause models. In all cases, it leads to weak persistent fluctuations of the low-density boundary and does not substantially deform the initial magnetopause profiles described by our analytical solutions with the particle distribution functions given by Eqs.~(\ref{eqDistribFunce4comp}),~(\ref{eqDistribFunci4comp}).

\begin{figure}[t]
\centering
    \includegraphics[width=0.6\linewidth]{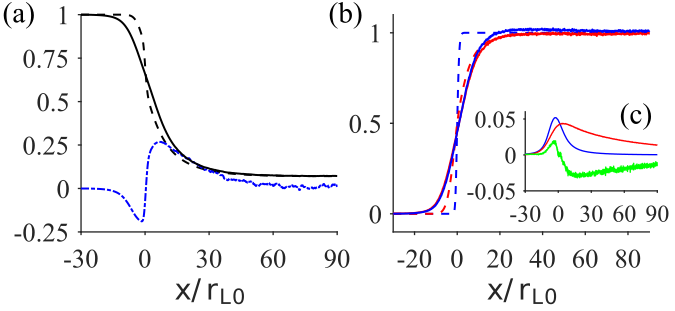}
	\caption{
        Structure of the two-component current sheet at the moment $t = 17500 \, \wplinv$:
        (a)~black solid line~--- magnetic field profile $B_z (x) / B_{z0}(-\infty)$, black dashes~--- its initial profile $B_{z0} (x) / B_{z0}(-\infty)$;
        blue~--- their relative difference $B_z / B_{z0} - 1$;
        (b)~red and blue solid lines~--- number densities of current-carrying ions and electrons, $n_\mathrm{i1}(x) / N_\mathrm{e1}$ and $n_\mathrm{e1}(x) / N_\mathrm{e1}$, respectively; dashes~--- their initial profiles;
        (c)~green~--- the difference of the ion and electron number densities $(n_\mathrm{i1} - n_\mathrm{e1}) / N_\mathrm{e1}$, proportional to the charge density of current-carrying particles; red and blue~--- number densities of the background ions and electrons, $n_\mathrm{i0}(x) / N_\mathrm{e1}$ and $n_\mathrm{e0}(x) / N_\mathrm{e1}$, respectively.
	}
	\label{fig3}
\end{figure}

Namely, in the case under consideration, according to Figs.~\ref{fig1}--\ref{fig3}, to the right of the low-density boundary ''plasma--vacuum'', the self-consistent magnetic field $B_z$ and plasma number density remain uniform along the~$y$ axis. The profiles of their dependence on the~$x$ coordinate only slowly flatten out, probably due to the influence of the electric field in the central part of the sheet arising due to thermal movement of the background particles and violation of charge compensation.
By the end of the simulation, the magnetic field profile $B_z(x)$ averaged along the~$y$ axis remains monotonic and the magnitude of its relative change, $|B_z (t) / B_{z0} - 1|$, where $B_{z0} = B_z(t=0, x)$, does not exceed 30\% (Fig.~\ref{fig3}(a)). It is noteworthy that the number density profiles of current-carrying plasma components (Fig.~\ref{fig3}(b),(c)) become equal almost everywhere, so that the charge density associated with them decreases drastically compared to the initial one (Fig.~\ref{fig1}(c)).

The same transient behavior, the absence of the Weibel-type instability and the development of weak bending-type instability of the magnetopause boundary are observed in similar two-component current sheets with Kappa energy distribution of particles. For brevity, we omit the analysis of the resulting persistent boundary fluctuations and postpone the description of examples of this kind to further publications. Thus, the two-component magnetopauses with particle distributions of the form (\ref{eqDistribFunce4comp}), (\ref{eqDistribFunci4comp}) in the absence of countercurrents are quite stable and remain close to the initial analytical solutions during rather long-term kinetic simulations.

\subsection{Four-component unstable current sheet with Kappa energy distributions of particles}
\label{sec:4cKappa}

\begin{figure}[b]
\centering
    \includegraphics[width=0.9\linewidth]{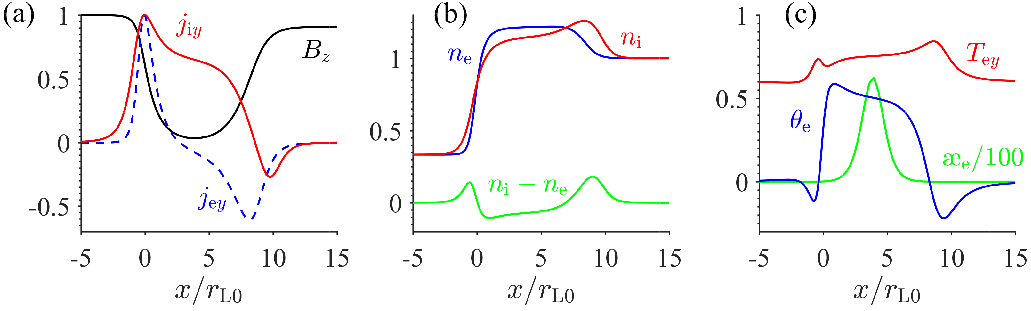}
	\caption{
        Structure of the four-component Kappa current sheet at the initial moment of time:
        (a)~black~--- magnetic field profile $B_z$, normalized to its maximum $B_z(-\infty)$;
        red and blue~--- current densities of ions and electrons respectively, normalized to their maxima;
        (b)~red~--- normalized number density of current-carrying ions, $(n_\mathrm{i1} + n_\mathrm{i2})/N_\mathrm{e1}$, blue~--- the same for current-carrying electrons $(n_\mathrm{e1} + n_\mathrm{e2})/N_\mathrm{e1}$, green~--- their difference, proportional to their charge density;
        (c)~blue~--- effective anisotropy $\theta_\mathrm{e}$ of all electrons including background ones, red~--- their effective temperature $\tilde{T}_y$, green~--- their joint demagnetization parameter $\varkappa_\mathrm{e}/100$.
	}
	\label{fig4}
\end{figure}

To demonstrate the possibility of development of the Weibel turbulence inside the sheet, we chose a model in which two pairs of electron and ion components with Kappa energy distributions $F_{\alpha s}(p)$, given by Eq.~(\ref{eqDistribFuncKappa}), form countercurrents. These distributions are sampled with an algorithm given in Ref.~\onlinecite{Zenitani2022} and had the parameter $\kappa = 3$. 
Electrons and ions forming currents in the direction of the $y$ axis, in the region of their isotropy to the right of the sheet, have effective temperatures of $T_\mathrm{e1} = 1500$~eV and $T_\mathrm{i1} = 300 $~eV and equal number densities $N_\mathrm{e1,i1} = 10^6$~cm$^{-3}$; those forming countercurrents, temperatures $T_\mathrm{e2} = 4500$~eV, $T_\mathrm{i2} = 300$~eV and equal number densities $N_\mathrm{e2,i2} = N_\mathrm{e1} / 3$ to the left of the sheet.
The countercurrents are shifted in space by adding constants $A_{\alpha 2}$ to the second pair of components in the distributions~(\ref{eqDistribFunce4comp})--(\ref{eqDistribFunci4comp}): $A_\mathrm{e2} = 2 (2 \me T_\mathrm{e1})^{1/2}$, $A_\mathrm{i2} = 2 (2 \mi T_\mathrm{i1})^{1/2}$, $A_\mathrm{e1,i1} = 0$.
To ensure the existence of a region with a weak magnetic field, we set the parameter $\tilde{P}_0 = 0.252$.
The background electrons and ions had a temperature of $T_\mathrm{e0,i0} = 1$~eV and number density profiles $n_\mathrm{e0}(x) = N_\mathrm{e1} - n_\mathrm{e1}(x) + 2 N_\mathrm{e2} - n_\mathrm{e2}(x)$, $n_\mathrm{i0} (x) = N_\mathrm{i1} - n_\mathrm{i1}(x) + 2 N_\mathrm{i2} - n_\mathrm{i2}(x)$. 
The structure of the sheet at the initial moment is shown in Fig.~\ref{fig4} (a part of the simulation domain is shown).

The gyroradius of a thermal electron at the conditional center of the sheet ($x = 0$) is $\rL \approx 990$~cm. In the region $1 < x / \rL < 7$, where the magnetic field is suppressed and the demagnetization parameter noticeably exceeds $1$, reaching a maximum of $\varkappa_\mathrm{e} (x /\rL \approx 4) \approx 60$, the effective anisotropy is $\theta_\mathrm{e} \approx 0.5$. 
Therefore, in the constructed sheet the Weibel perturbations with characteristic wavelengths $\sqrt{3} \lambdam \approx 15 \, c/\wpl$ can grow effectively at time scales of the order of $\Gamma^{-1} \sim 380 \, \wplinv$.
Due to limited computational resources, a small two-dimensional region with dimensions $L_x = 60\, c/\wpl \approx 34\, \rL$ and $L_y = 96\, c/\wpl \approx 9\, \lambdam$ is used, divided by $2500 \times 1000$ cells.
The boundary conditions at $x = -0.275 \, L_x$ and $x = 0.725 \, L_x$ are reflective for particles and absorbing for fields, the boundary conditions at $y = 0$ and $y = L_y$ are periodic. The four current-carrying plasma components comprise approximately $8 \times 10^8$ macroparticles, and the two background components comprise approximately $3 \times 10^8$ ones.

\begin{figure}[b]
\centering
    \includegraphics[width=0.9\linewidth]{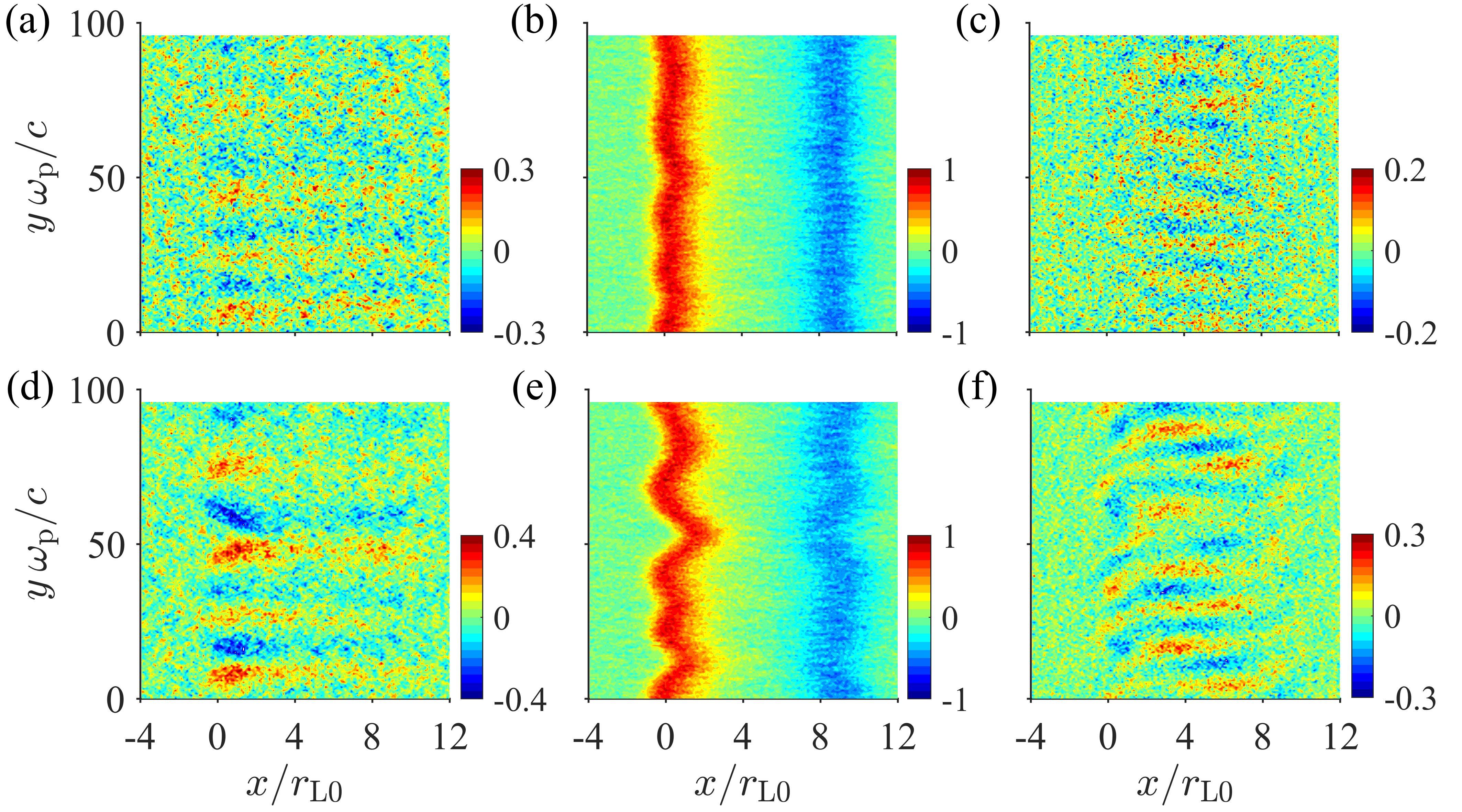}
	\caption{
        Total current density in the Kappa sheet with $\tilde{P}_0 = 0.252$ at the moments (a)--(c)~$t \approx 4600 \, \wplinv$, (d)--(f)~$t \approx 6900 \, \wplinv$. Panel columns correspond to projections $j_x$, $j_y$, $j_z$ (from left to right) normalized to $j_\mathrm{e1}(t = 0, x = 0)$. 
        }
	\label{fig5}
\end{figure}

Snapshots from the simulation of the sheet's evolution are shown in Fig.~\ref{fig5}. 
Starting from the moment $t \approx 3400 \, \wplinv$, modulation in the $y$-direction appears in the two-dimensional distribution of the magnetic field $B_z$. These perturbations are maintained by the system of in-plane currents clearly seen in the distributions of the main current density component $j_y$ (as perturbations of its initial profile) and the component $j_x$.
The component $j_z$ also begins to grow from the noise and demonstrates a similar modulation.

\begin{figure}[t]
\centering
    \includegraphics[width=0.35\linewidth]{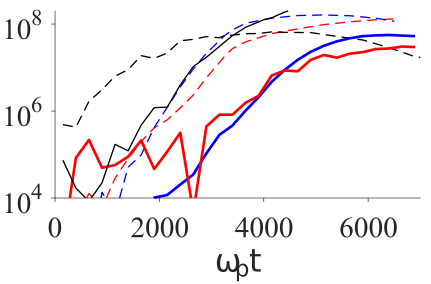}
	\caption{
        Evolution of the fastest growing modes of the current density components $j_x$~(solid red line) and $j_z$~(solid blue) in the Kappa sheet with $\tilde{P}_0 = 0.252$. Each mode's power is shown in a.u.
        Spatial spectra along the axis $y$ are obtained after averaging both components over the interval $2 < x / \rL < 7$.
        Black line corresponds to the fastest mode of $j_x$ in the additional simulation with $\tilde{P}_0 = 0.282$, i.e., a stronger field $B_z$, where $j_z$ currents do not emerge.
        Dashed lines are for the Maxwellian sheet with the same parameters $N_{\alpha s}$, $T_{\alpha s}$, but $\tilde{P}_0 = 0.282$ and $0.56$ for the cases of weak and strong field, respectively.
	}
	\label{fig6}
\end{figure}

In the spatial spectra of the components $j_x$ and $j_z$ along the axis $y$, averaged over the interval $2 < x / \rL < 7$, a stage of exponential growth of the lower modes is visible, continuing until saturation at the moment $t \approx 4600 \, \wplinv$ (Fig.~\ref{fig6}). Note that the averaging interval is chosen to conveniently obtain the spectral properties of the current perturbations, but does not include the region $0 < x / \rL < 2$, where the in-plane perturbations are strongest (see Fig.~\ref{fig5}). Thus, the level of saturation of the latter is higher than Fig.~\ref{fig6} implies.
The fastest growing modes are $5$ and $8$ respectively and have dimensionless wavelengths $\lambda \wpl / c \approx 19$ and $12$. 
Their inverse growth rates are $\sim \! 1100 \wplinv$ and $\sim \! 720 \wplinv$, quite higher than the analytical prediction of $380 \, \wplinv$ for the Weibel instability, probably due to inaccuracy of the bi-Maxwellian approximation used. 
At later stages we see no evidence of self-similar power-law behavior typical of the Weibel instability spectral evolution~\cite{Borodachev2017, Silva2020, Zhou2021, Nechaev23_JPP}, perhaps due to insufficient time elapsed, the presence of the sheet's magnetic field and a small transverse size $L_y$ of the simulation domain.

\begin{figure}[t]
\centering
    \includegraphics[width=0.3\linewidth]{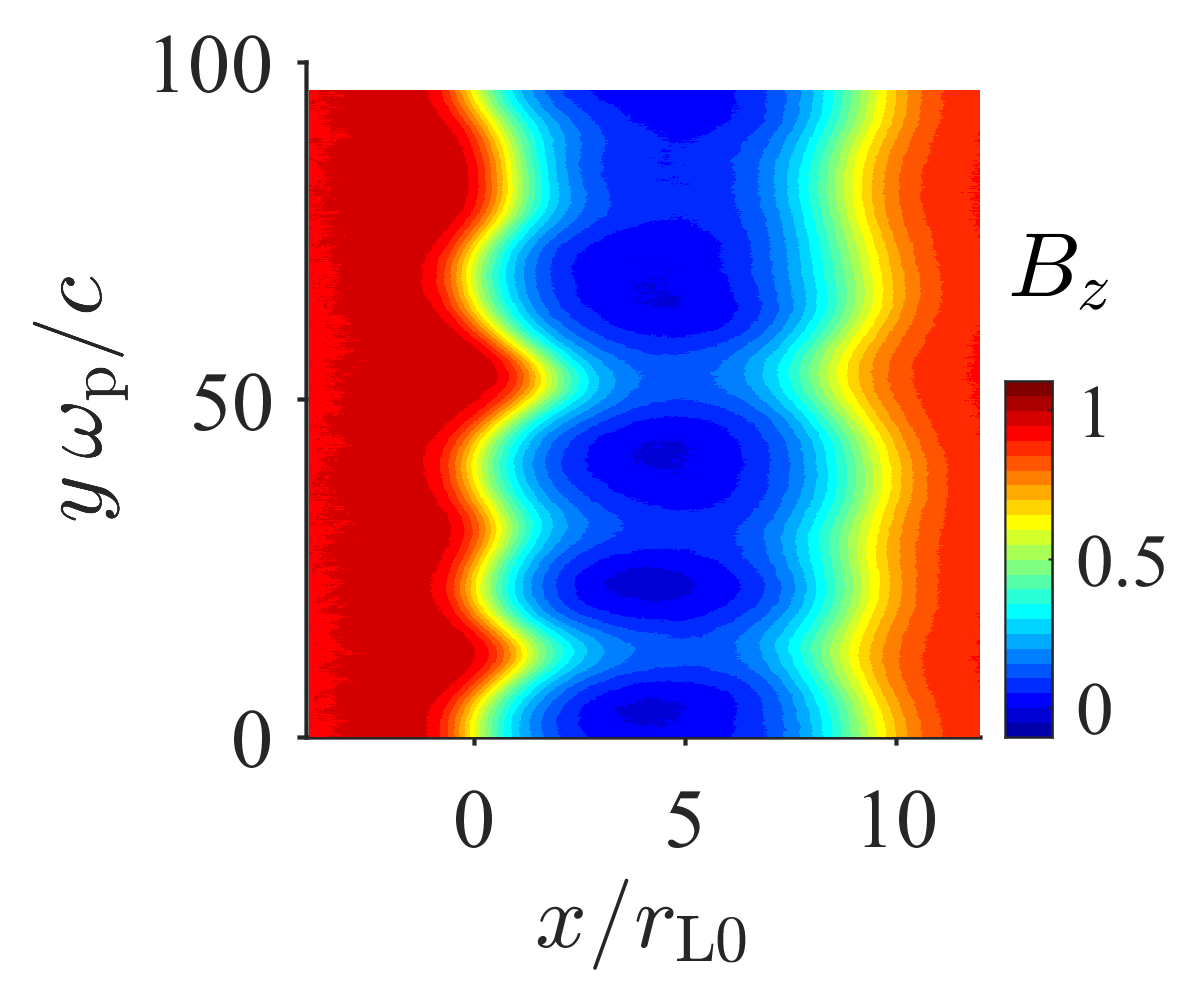}
	\caption{
        Magnetic field $B_z$ normalized to its maximum initial value in the Kappa sheet with $\tilde{P}_0 = 0.252$ at the moment $t \approx 6500 \, \wplinv$.
	}
	\label{fig7}
\end{figure}

The simulation stopped at the moment $t \approx 6900 \, \wplinv$. By that time the joint effective anisotropy of all electrons near the magnetic field's minimum is $\theta_\mathrm{e} \approx 0.25$. 
Note that despite the magnetic turbulence developing inside, the overall structure of the sheet remains stable (Fig.~\ref{fig7}).

\begin{figure}[b]
\centering
    \includegraphics[width=0.55\linewidth]{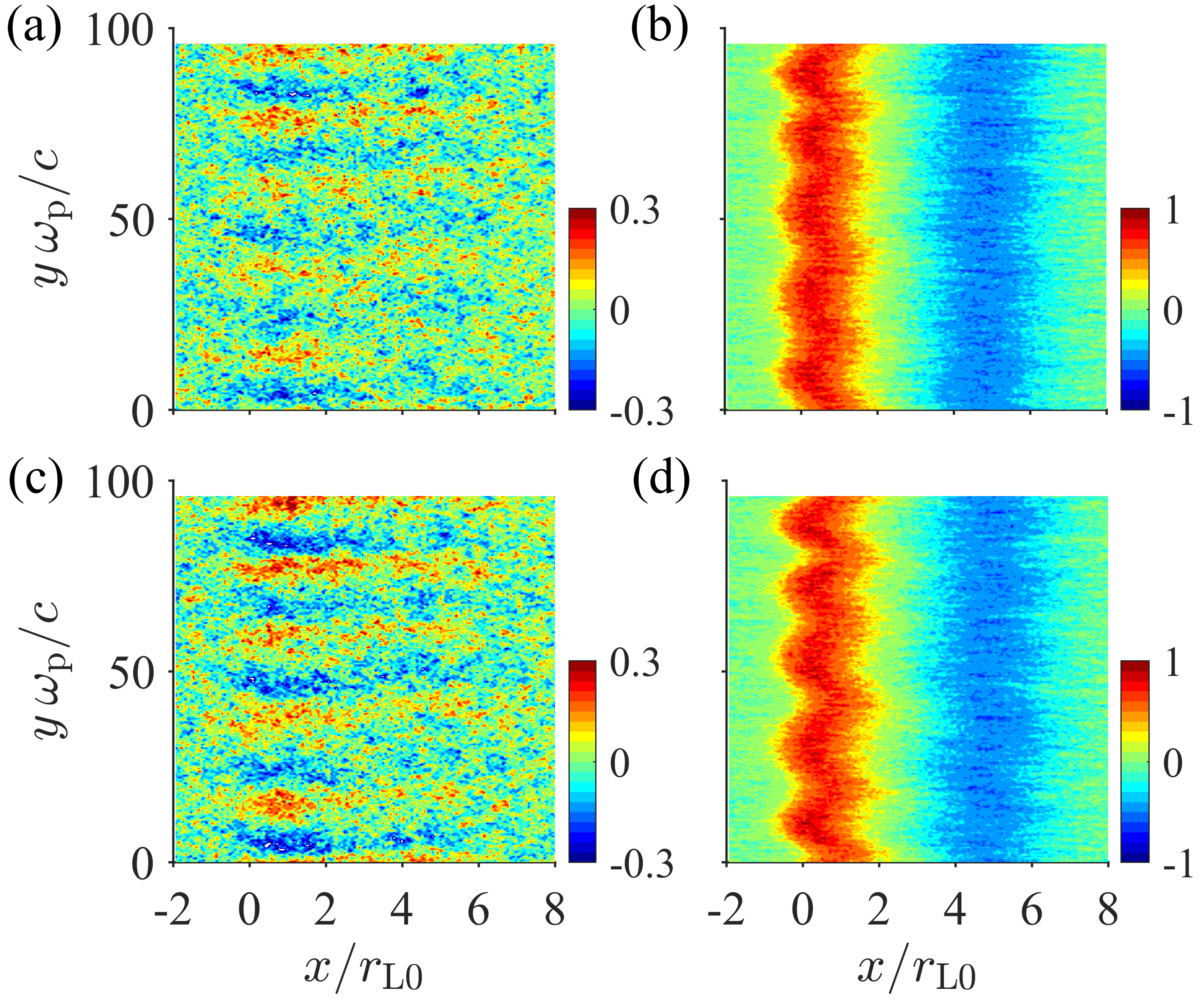}
	\caption{
        Total current density in the Kappa sheet with $\tilde{P}_0 = 0.282$ at the moments (a)--(b)~$t \approx 3400 \, \wplinv$, (c)--(d)~$t \approx 4000 \, \wplinv$. Panel columns correspond to projections $j_x$, $j_y$ (from left to right) normalized to $j_\mathrm{e1}(t = 0, x = 0)$. 
        The $j_z$ projection contains only noise, as the Weibel instability is inhibited.
        }
	\label{fig8}
\end{figure}

To verify the theoretical conclusions of Section~\ref{sec:theory:stability}, we carry out an additional simulation with an increased parameter $\tilde{P}_0 = 0.282$. Consequently, the initial magnetic field $B_z$, given by Eq.~(\ref{eqBgeneral}), increases $6.5$ times in its minimum, while the anisotropy near this minimum remaines approximately the same, $\theta_\mathrm{e}(x / \rL \approx 2) \approx 0.5$. Hence, the electrons in the sheet are almost completely magnetized with the parameter $\varkappa_\mathrm{e} < 1.5$. 
As follows from the simulation results, indeed, a stronger field prevents the currents $j_z$ from growing above the noise level for the entire simulation duration of $4000 \, \wplinv$ (Fig.~\ref{fig8}). 
In contrast, the currents in the $xy$~plane still emerge, and their growth is even faster. 
This is illustrated in Fig.~\ref{fig6} with the black line corresponding to the fastest growing mode of the $j_x$ current component.
Apart from that, the distribution of the in-plane currents $j_x$, $j_y$ is similar to the one presented in Fig.~\ref{fig5} (first two columns) for the case of a weak field $B_z$. 
The existence of their perturbations in such a strong field indicates that this instability is not of a Weibel type. It is likely the bending-type instability observed in the two-component sheet of Section~\ref{sec:2cMaxw}.

On the whole, the considered current sheets, either with or without small-scale turbulence inside, demonstrate the long-term structural stability even in the presence of countercurrents of different particle components and bending deformation of the sharp low-density boundary of the magnetopause model.

\subsection{Four-component unstable current sheet with Maxwellian energy distributions of particles}
\label{sec:4cMaxw}

\begin{figure}[b]
\centering
    \includegraphics[width=0.9\linewidth]{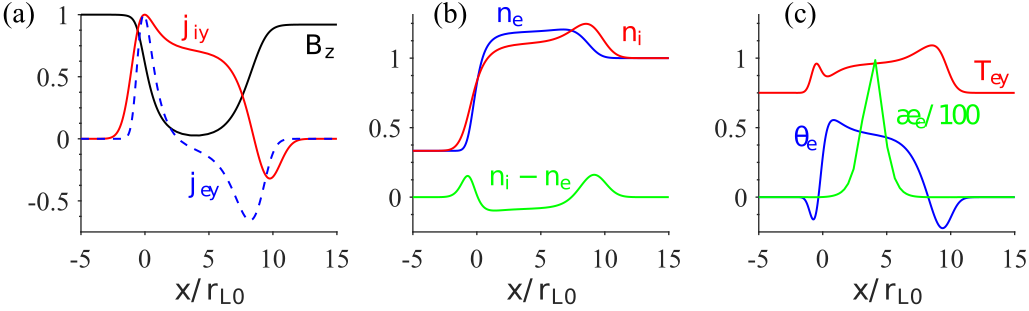}    
	\caption{
        Structure of the four-component Maxwellian current sheet at the initial moment of time:
        (a)~black~--- magnetic field profile $B_z$, normalized to its maximum $B_z(-\infty)$;
        red and blue~--- current densities of ions and electrons respectively, normalized to their maxima;
        (b)~red~--- normalized number density of current-carrying ions, $(n_\mathrm{i1} + n_\mathrm{i2})/N_\mathrm{e1}$, blue~--- the same for current-carrying electrons $(n_\mathrm{e1} + n_\mathrm{e2})/N_\mathrm{e1}$, green~--- their difference, proportional to their charge density;
        (c)~blue~--- effective anisotropy $\theta_\mathrm{e}$ of all electrons including background ones, red~--- their effective temperature $\tilde{T}_y$, green~--- their joint demagnetization parameter $\varkappa_\mathrm{e} / 100$.
	}
	\label{fig9}
\end{figure}

In order to clarify the dependence of the instability features on the type of particle energy distribution, we construct a sheet with the same plasma parameters as in Section~\ref{sec:4cKappa}, but with a Maxwellian distribution, which is the limit of the Kappa distribution with $\kappa \to \infty$. We set the parameter $\tilde{P}_0 = 0.282$. 
The structure of the sheet at the initial moment (Fig.~\ref{fig9}) qualitatively resembles that of the Kappa sheet with $\kappa =3$, but the current and number densities of individual plasma components have a bit smaller characteristic scales, which is due to heavier energy tails of the Kappa distribution compared to the Maxwellian~\cite{Kocharovsky2021}.

Simulations of the evolution of this current sheet as well as other complex sheets of the same class with Maxwellian particle energy distributions confirm the conclusions of Section~\ref{sec:4cKappa}. Namely, it is shown that a certain choice of particle distributions, provided they are smooth and qualitatively similar to the Kappa distribution with the parameter $\kappa \sim 3$, does not change, in comparison with the latter, neither the global structure of the current sheet, nor the character of the Weibel-type instability in the inner part of the sheet with a weak regular magnetic field. Hence, the possibility of long-term existence of small-scale quasi-magnetostatic turbulence in distributed current sheets seems to be quite universal in the presence of countercurrents formed by different particle components at the opposite sides of a complex magnetopause. The main general dimensionless condition for the Weibel-type instability development and formation of quasi-magnetostatic turbulence in the four-component magnetopause models is the fulfillment of the inequality on demagnetization, $\varkappa_\mathrm{e} > 1$, which implies a sufficiently high effective anisotropy parameter (not much smaller than unity in realistic PIC-simulations).

These conditions could be met for the parameters of the Earth's magnetopause, if it was of a four-component type with separated countercurrents and a weak enough self-consistent magnetic field between them.  Similarly, using the above dimensionless conditions required for the existence of Weibel turbulence, one can estimate whether the turbulent four-component magnetopauses are possible for the typical plasma parameters of other cosmic objects, for example, exoplanets or stars moving in a magnetized interstellar medium or supporting a magnetized stellar wind. 

According to the present work, if these conditions are not fulfilled, e.g., in the case of a rather strong internal magnetic field in a thin four-component magnetopause, then the short-wavelength Weibel-type instability does not occur in both cases of Kappa and Maxwellian distributions of particles. For example, if in the magnetopause model simulated both in this Subsection for the Maxwellian distribution and in the previous Subsection for the Kappa distribution we take $\tilde{P}_0 = 0.282$ instead of $\tilde{P}_0 = 0.252$, then the Weibel-type instability disappears inside the sheet, i.e., in the internal region between the countercurrents. At the same time, the bending-type instability at the sharp low-density boundary of a complex current sheet is established in all cases under considaration and makes the whole sheet a bit thicker and smoother, though does not qualitatively change its global structure.

\section{Conclusion}
\label{sec:concl}

In this work, we investigate the development of small-scale kinetic instabilities in two- and four-component current sheet equilibria, constructed analytically and representing a wide class of models of a distributed magnetopause with arbitrary particle energy distributions~\cite{Nechaev23_JETPLen}. 
With a suitable choice of parameters, these models can embed a considerably wide region, where the magnetic field is weak and the collisionless plasma has a significant self-consistent anisotropy of the particle velocity distribution. 
As a result, according to two-dimensional (2D3V) PIC-simulations with the code EPOCH for the cases of Kappa and Maxwellian energy distributions, we show that, depending on the model parameters, the Weibel-type instability can be inhibited or develop effectively, leading to quasi-magnetostatic turbulence inside the distributed sheet. Both the stability conditions and the properties of this instability, including the saturation level of small-scale turbulence, weakly depend on the certain choice of the particle energy distribution in the considered range of the Kappa parameter, $\kappa \geq 3$ (Maxwellian distribution corresponds to $\kappa \to \infty$). The latter is also valid for the bending-type instability at the sharp edge of a magnetopause, which occurs in all simulations and leads to small-scale fluctuations of the plasma density and magnetic field.

We identify the features of emergence and nonlinear evolution of quasi-magnetostatic Weibel turbulence in the presence of a nonuniform self-consistent magnetic field and an inhomogeneous plasma with particle-velocity anisotropy. 
We show that, despite the small-scale bending-type perturbations of the initial particle currents on the sharp low-density edge and the turbulent magnetic fields, if any, created by the Weibel-type instability in the inner part of the complex sheet, the large-scale structure of the magnetopause model does not change notably and remains quasi-stable during the time period achievable in our simulations, including the stage of saturation of turbulence.

The proposed current sheet models and the established properties of their small-scale instabilities, leading to internal turbulence and boundary fluctuations, can be useful for the analysis and interpretation of the physical phenomena in the vicinity of the outer parts of planetary magnetospheres exposed to stellar winds and the coronas of late-type stars. For these applications, PIC-simulations of a large set of suggested multicomponent models of a complex magnetopause, admitting countercurrents and persistent turbulence, should be carried out in a wide range of their parameters. These simulations should include the suggested more general models of a distributed magnetopause with a shear of magnetic field lines~\cite{Nechaev23_JETPLen, Kocharovsky2016_UFN}, for such configurations are typical for the magnetized wind of late-type stars and have been observed in the solar wind (see, e.g., Refs.~\onlinecite{Louarn2004, Dunlop2008, Norgren2018, Lotekar2022, Vasko2022, Vasko2024}). There are also significant open problems related to long-term evolution of both the Weibel-type and bending-type turbulence, as well as persistent support of such turbulence by some kind of energetic particle injection into a magnetopause. Generally, solutions to these problems will go beyond the simple models of the present work, require 3D PIC modeling of nonplanar current sheets with nonzero plasma flows, and involve more types of turbulence, including plasma-oscillations (Langmuir-waves), whistler and lower-hybrid drift turbulence observed in the Earth's magnetopause and magnetosheath.

\begin{acknowledgments}
The work was supported by the Russian Science Foundation grant No.~20-12-00268. Computational resources were provided by the Joint Supercomputer Center of~the Russian Academy of~Sciences.
\end{acknowledgments}

\bibliography{biblio}

\end{document}